\documentclass[conference]{IEEEtran}
\IEEEoverridecommandlockouts
\usepackage[
    natbib=true,
    style=numeric,
    sorting=none,
    maxnames=1
]{biblatex} 
\usepackage{amsmath,amssymb,amsfonts}
\usepackage{algorithmic}
\usepackage{graphicx}
\usepackage{textcomp}
\usepackage{caption}
\usepackage{subcaption}
\usepackage{multirow}
\usepackage{xcolor}
\def\BibTeX{{\rm B\kern-.05em{\sc i\kern-.025em b}\kern-.08em
    T\kern-.1667em\lower.7ex\hbox{E}\kern-.125emX}}

\begin{document}

\title{Error assessment of microwave holography inversion for shallow buried objects\\

}

\author{\IEEEauthorblockN{1\textsuperscript{st} Emanuele Vivoli}
\IEEEauthorblockA{\textit{MICC} \\
\textit{Università degli Studi di Firenze}\\
Florence, Italy \\
emanuele.vivoli@unifi.it}
\and
\IEEEauthorblockN{2\textsuperscript{nd} Luca Bossi}
\IEEEauthorblockA{\textit{Department of Information Engneering} \\
\textit{Università degli Studi di Firenze}\\
Florence, Italy \\
l.bossi@unifi.it }
\and
\IEEEauthorblockN{3\textsuperscript{rd} Marco Bertini}
\IEEEauthorblockA{\textit{MICC} \\
\textit{Università degli Studi di Firenze}\\
Florence, Italy \\
marco.bertini@unifi.it}
\and
\IEEEauthorblockN{4\textsuperscript{th} Pierluigi Falorni}
\IEEEauthorblockA{\textit{Department of Information Engneering} \\
\textit{Università degli Studi di Firenze}\\
Florence, Italy \\
pfalorni@gmail.com}
\and

\IEEEauthorblockN{5\textsuperscript{th} Lorenzo Capineri}
\IEEEauthorblockA{\textit{Department of Information Engneering} \\
\textit{Università degli Studi di Firenze}\\
Florence, Italy \\
lorenzo.capineri@unifi.it}

}
\maketitle

\begin{abstract}
Holographic imaging is a technique that uses microwave energy to create a three-dimensional image of an object or scene. This technology has potential applications in land mine detection, as the long-wavelength microwave energy can penetrate the ground and create an image of hidden objects without the need for direct physical contact. However, the inversion algorithms commonly used to digitally reconstruct 3D images from holographic images, such as Convolution, Angular Spectrum and Fresnel, are known to have limitations and can introduce errors in the reconstructed image. Despite these challenges, the use of holographic radar at around 2 GHz in combination with holographic imaging techniques for land mine detection allows to recover size and shape of buried objects.
In this paper we estimate the reconstruction error for the convolution algorithm based on hologram imaging simulation, and assess these errors recommending an increase in the scanner area, considering the limitations that the system has and the expected error reduction.
\end{abstract}


\section{Introduction}

Holographic image reconstruction is arguably the most effective processing algorithm for microwave land mine detection. Holography makes it possible to gather both amplitude and phase data and can produce both 2-D and 3-D images of objects. The image-reconstruction computer algorithms examined in this research were originally derived from microwave holography techniques \cite{microwave:holo}. 
Both acoustic \cite{acustic:holo} and microwave holography \cite{6254884} are long-wavelength implementations of the original optical holography techniques created by Gabor et al. \cite{gabor:1948} in the 1940s, and they are remarkably similar each other.

All three forms of holographic imaging operate by sampling the amplitude and phase of a wavefront scattered from a target object.
The sampled wavefront is subsequently ``reconstructed`` either optically or using computer image-reconstruction methods that are based on Fourier optics.

The commonly used reconstruction algorithms are based on the Fresnel equation or convolution methods \cite{942570}. The methods based on convolution are suitable for application where the scattering object is in the near field region, because is not required the parallax approximation. Moreover, this approach is suitable for synthetic aperture radar (SAR) imaging  \cite{134436, 128031, 10.1007/978-1-4615-8210-6_18}.

\begin{figure}
\vspace{-0.6cm}
\centering
\includegraphics[scale=0.25]{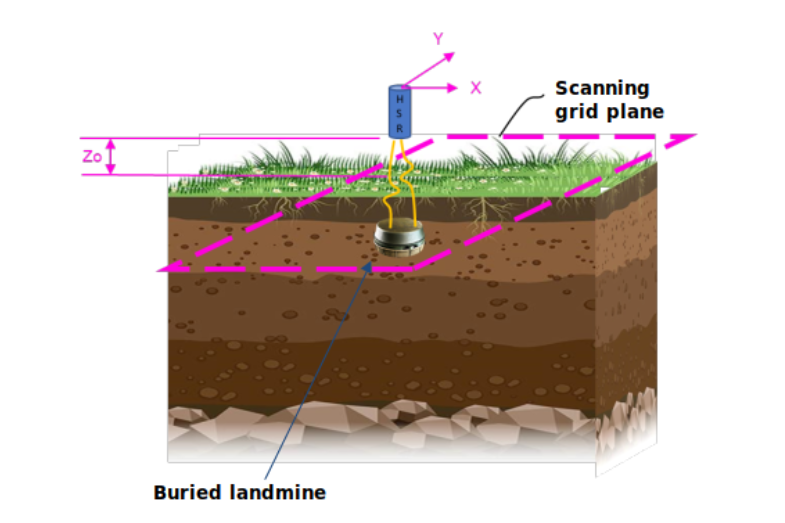}
\caption{Illustration of the geometry of holographic acquisition. The dashed line define the scanning plane. The HSR operates in CW @ 2 GHz. \cite{electronics11182883}}
\label{fig:geometry}
\vspace{-6mm}
\end{figure}

\begin{figure}
    \centering
    \includegraphics[scale=0.20]{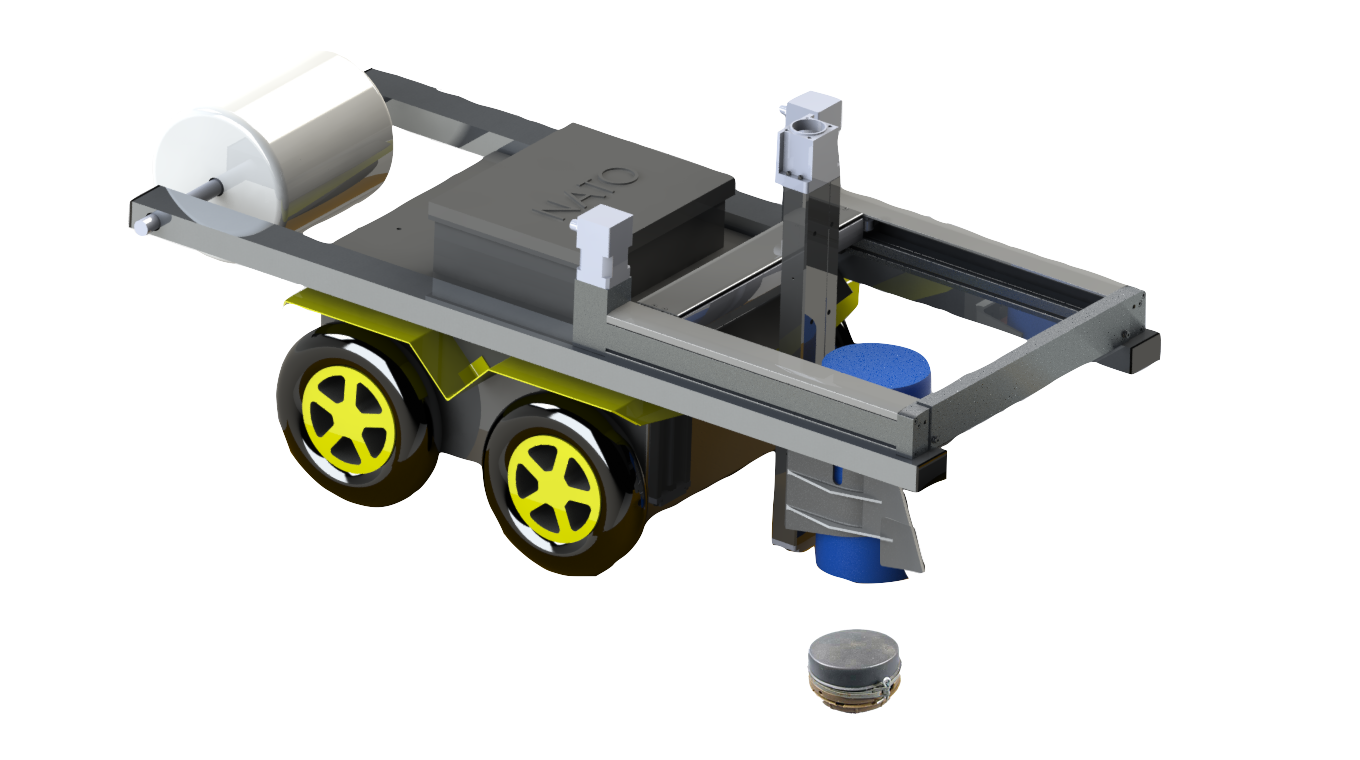}
    \caption{CAD rendering of Robot "UGO-1st", scanning PMN-4 at distance from HSR radar.}
    \label{fig:ugo-first}
\end{figure}

The HSR system (Holographic Subsurface Radar) uses a circular waveguide antenna with a single feed. A circulator device is employed for separating the transmitted and the received signals. The hologram is formed by the interference between the transmitted signal and the received signal from the antenna. 
The HSR system is mounted on a fast mechanical scanner that is capable of operating outdoors, providing uniform spatial sampling of 5mm by 5mm over an area as large as 300 mm x 170 mm. In Fig. \ref{fig:ugo-first}, are represented the robot with the mechanical scan and HSR system on it. As part of the NATO SPS G5731 project \cite{nato:demining-robot}, the robot platform carrying several sensors imply some constraint on the dimensions of the scanned area.
The digitally-recorded microwave holograms contain detailed information about the size and shape of buried objects within the effective signal penetration depth.

\section{Theory}
The algorithm used in this paper is inspired from \cite{942570}, where the system configuration is represented in Fig. \ref{fig:geometry}, in which we can see that the source is located at $(X, Y, Z_0)$ and a general point in the target located at $(x, y, z)$. The target is characterized by a reflective function $f(x, y, z)$ which lead to the definition of the reflection perceived at the transceiver calculated by the superposition of each point on the target. The image reconstruction is numerically calculated by the following formula:
\begin{equation*}
    f(x,y,z) = FT^{-1}_{2D}\big\{ FT_{2D}\{s(x,y,z)\}e^{-j z \sqrt{4k^2-k_x^2-k_y^2}}\big\}
\end{equation*}
in which $FT$ represents the Fourier Transform operation, and $s(x,y,z)$ is the response measured at the transceiver. 

Following the above formulation, many tools have attempted to numerically solve the image reconstruction by proposing various implementation of the algorithm in different programming languages. Previous work \cite{holo:book} has defined the MATLAB algorithms of Angular Spectrum and Convolution reconstruction. More recent tools \cite{delafuente, holopy} adopted Python language in structuring an open source tool that is able to work with digital holograms and light scattering. In particular, which has been our choice for experimentation, HoloPy \cite{holopy} is able to do so by forward propagation of light from a scattering calculation of a predetermined scatterer, as well as backward propagation of light from a digital hologram to reconstruct slices of 3D volumes. 

\subsection{Optical holography}\label{sec:optic-holo}
To assess the reliability and performances of the chosen Python library \cite{holopy} we implemented a simulation composed by a polystyrene sphere object (index of refraction $n=1.59$) with radius $r=0.5$\;$ \mu m$, located at $c=(5, 5, 5)$\;$\mu m$. The complete setup is defined by the sphere in air (medium index $i_m= 1$), illuminated by red light (wavelength $w = 660$\;$ nm$ ) with polarization along the $x$ axis $pol\_axis = (1, 0)$. The detector is defined as a $100 \times 100$ pixel array, with each square pixel of side length $0.1$\;$ \mu m$.

\begin{figure}
    \centering
    \includegraphics[scale=0.35]{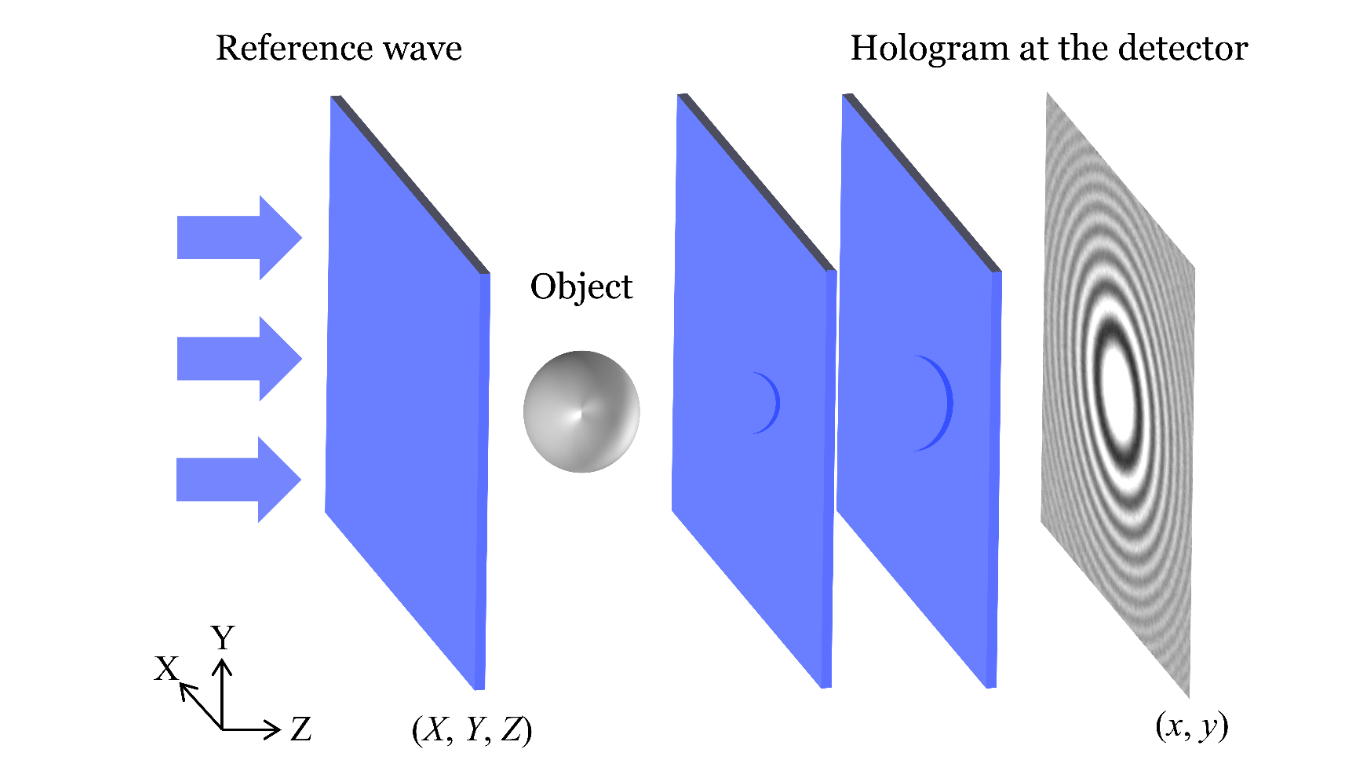}
    \caption{Schematic of in-line digital holography}
    \label{holo:scheme}
    \vspace{-5mm}
\end{figure}

\begin{figure}
\centering
\begin{subfigure}{.25\textwidth}
    \centering
    \includegraphics[scale=0.3]{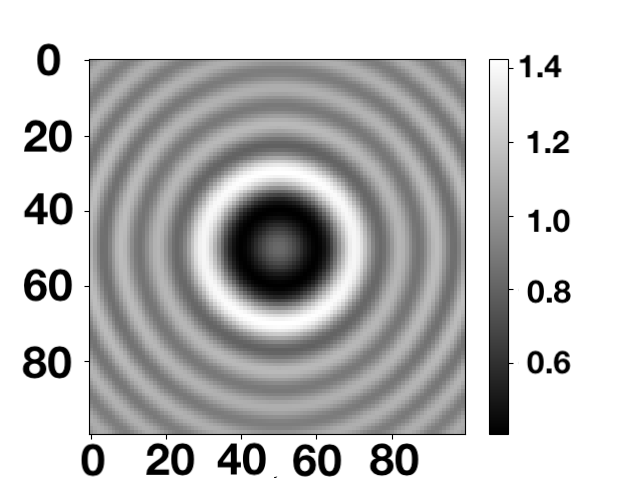}
    \caption{Simulated Hologram image.}
    \label{holo:sphere-optic}
\end{subfigure}%
\begin{subfigure}{.25\textwidth}
    \centering
    \includegraphics[scale=0.3]{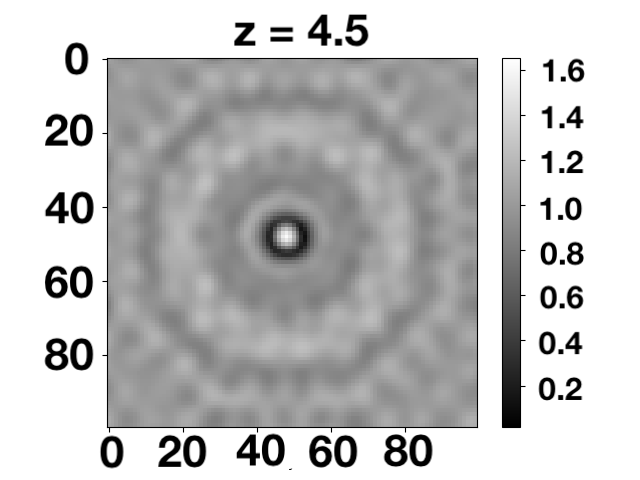}
    \caption{Reconstruction image.}
    \label{holo:recon-5}
\end{subfigure}
\caption{Simulated Hologram with Optical holography experimental setup (\ref{holo:sphere-optic}), and reconstructed image of hologram at distance $z=5$\;$ \mu m$ (\ref{holo:recon-5}). $x$ and $y$ axis are reported in $nm$.}
\vspace{-3mm}
\label{fig:holo-lossless}
\end{figure}

Once the light is forward propagated through the micro-sphere, the in-line hologram generated by a plane wave scattering is obtained (as in Fig. \ref{holo:scheme}). 
The process now is to reconstruct, with the HoloPy inversion algorithm implementation, the original 3D experimental setup in which the micro-sphere is located at position $(5, 5, 5)$\;$ \mu m$ in the simulated space. In so doing, we expect at distance $z=5$\;$ \mu m$, in the reconstructed image, to see a spherical object with radius $r=0.5$\;$ \mu m$. In fact, as shown in Fig. \ref{holo:recon-5}, theory is confirmed and in the obtained image there is the presence of a spherical object having the expected radius size.

\subsection{Holography for microwave imaging}\label{sec:holo-micro}

In the previous example we used wavelength and objects dimension related to Optical holography. However, as our first goal is to use these techniques in real application of landmine detection, the physics of the simulation is different. For this second example we use the same polystyrene sphere object (index of refraction $n=1.59$) with, this time, radius $r=5$\;$cm$, located at $c=(15, 15, 15)$\;$cm$. The sphere is immersed in air (medium index $i_m= 1$), illuminated by a microwave (wavelength $\lambda_w = 15$\;$cm$ \@f=2 GHz in air) with the same polarization as before $($along the $x$ axis $pol\_axis = (1, 0))$. The detector is defined as a $52 \times 62$ pixel array (as in \cite{nato:ugo-1st}), with each square pixel of side length $0.5$\;$cm$. The reconstructed image is shown in Fig. \ref{holo:sphere-real-recon}, in which we can see the low accuracy in reconstruction.

\begin{figure}
\centering
\begin{subfigure}{.25\textwidth}
    \centering
    \includegraphics[scale=0.3]{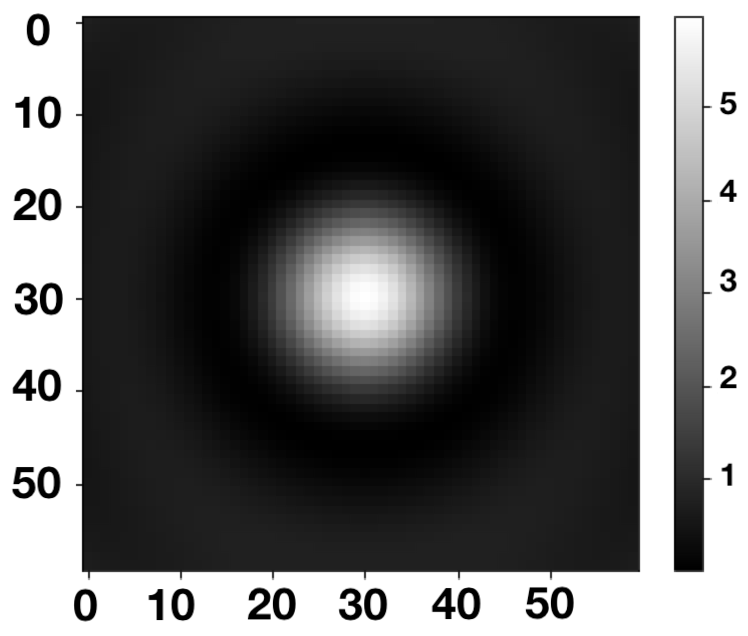}
    \caption{Simulated Hologram image.}
    \label{holo:sphere-real}
\end{subfigure}%
\begin{subfigure}{.25\textwidth}
    \centering
    \includegraphics[scale=0.3]{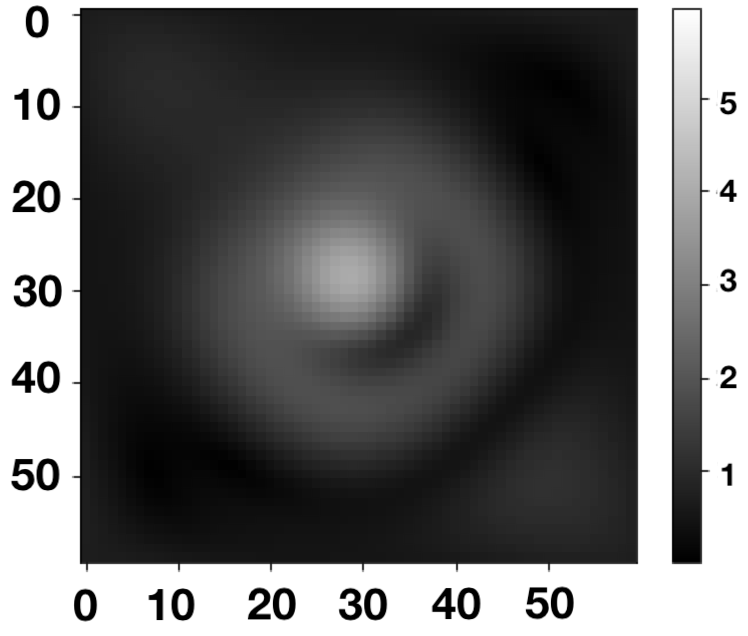}
    \caption{Reconstructed image.}
    \label{holo:sphere-real-recon}
\end{subfigure}
\caption{Simulated Hologram with realistic experimental setup as "UGO-1st" (\ref{holo:sphere-real}) \cite{9138227}, and reconstructed image of hologram at distance $z=5$\;$ cm$ (\ref{holo:sphere-real-recon}). $x$ and $y$ axis are reported in $cm$.}
\vspace{-3mm}
\label{fig:holo-lossy}
\end{figure}

\begin{figure}
\centering
\begin{subfigure}{.25\textwidth}
    \centering
    \includegraphics[scale=0.14]{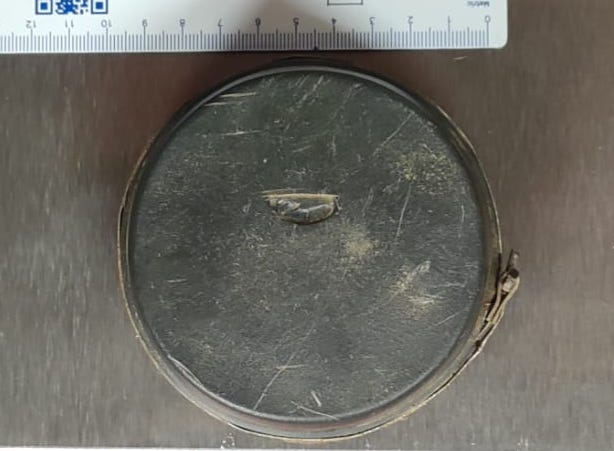}
    \caption{PMN4 Landmine sizes (d=$9.5$\;$cm$, h=$4.5$\;$cm$).}
    \label{holo:sphere-real}
\end{subfigure}%
\begin{subfigure}{.25\textwidth}
    \centering
    \includegraphics[scale=0.10]{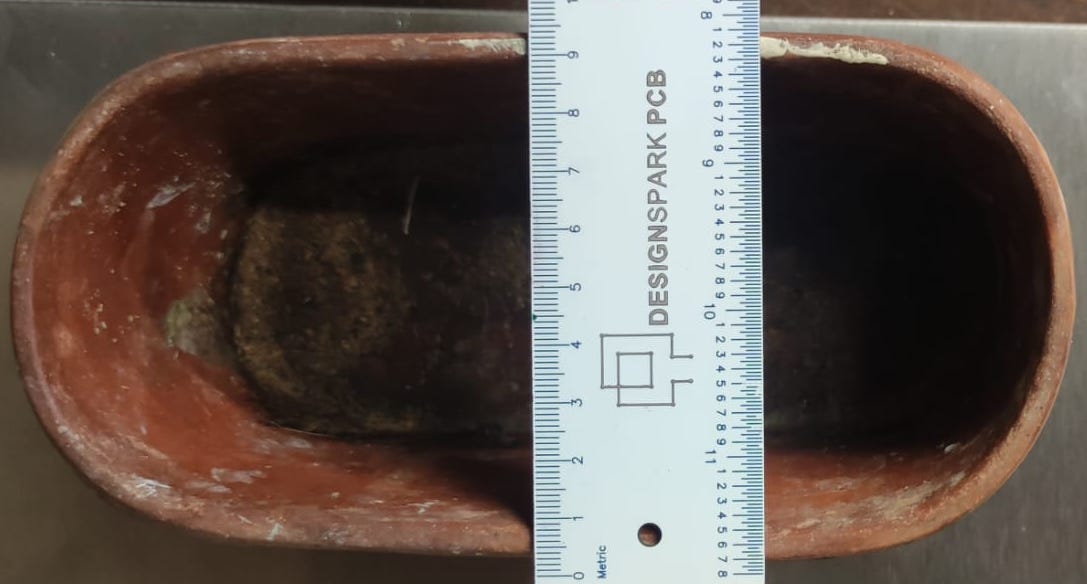}
    \caption{Terracotta vessel sizes (w=$9.5$\;$cm$, l=$19$\;$cm$ h=$8.5$\;$cm$).}
    \label{holo:sphere-real-recon}
\end{subfigure}
\caption{Pictures reporting dimensions of considered objects.}
\vspace{-4mm}
\label{fig:holo-lossy}
\end{figure}

\begin{figure}

\begin{tabular}{cc}

\subfloat[Hologram image (PMN4 landmine).]{ 
    \label{holo:pmn4}\includegraphics[width=0.2\textwidth]{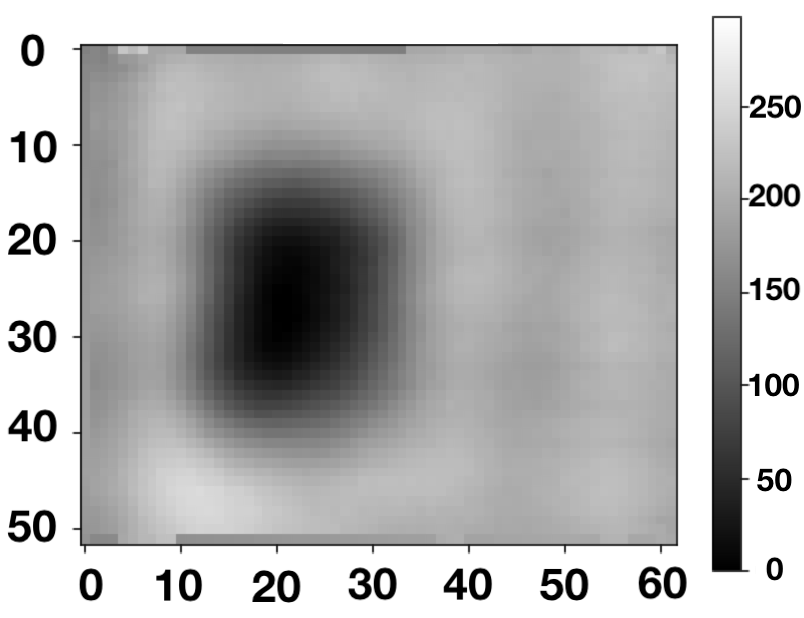}
}
&
\subfloat[Reconstructed image (PMN4 landmine).]{
    \label{holo:PMN4-recon}\includegraphics[width=0.2\textwidth]{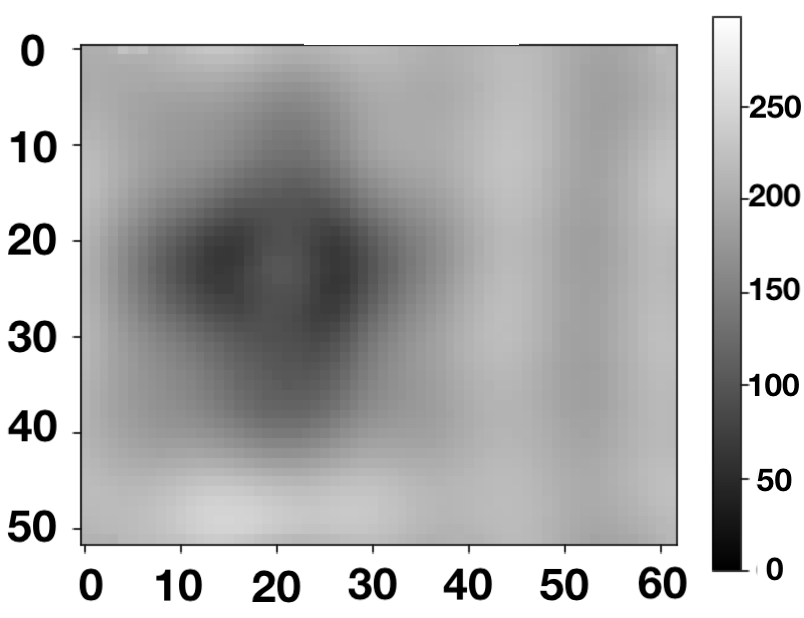}
}\\
\subfloat[Hologram image (Terracotta vessel).]{
    \label{holo:vassel}\includegraphics[width=0.2\textwidth]{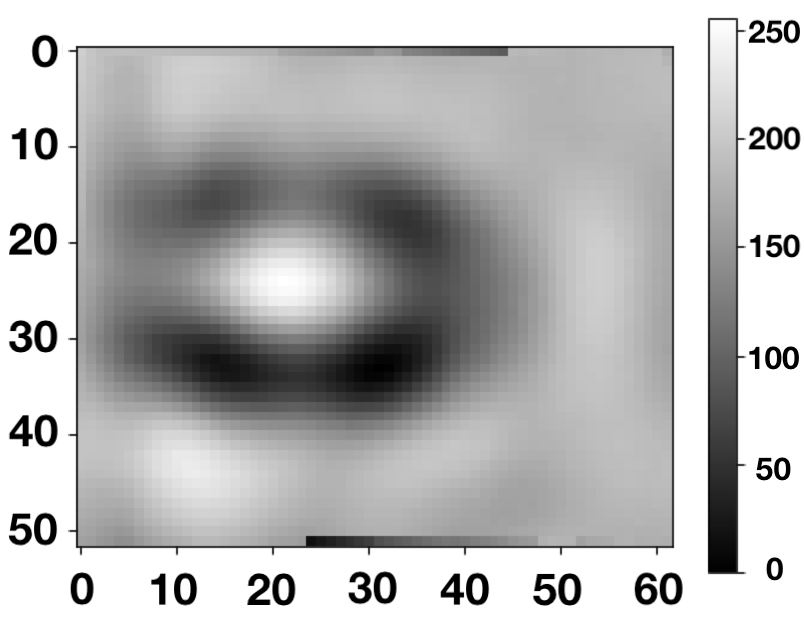}
}
&
\subfloat[Reconstructed image (Terracotta vessel).]{
    \label{holo:vassel-recon}\includegraphics[width=0.25\textwidth]{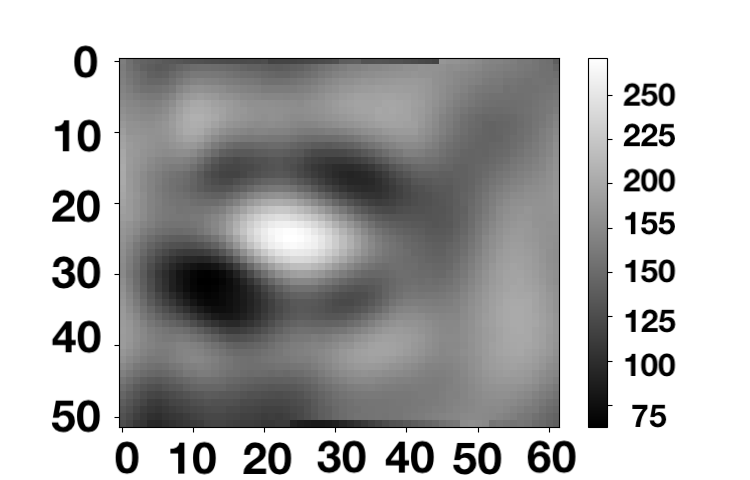}
} \\
\vspace{-5mm}
\end{tabular}

\caption{Hologram image taken by "UGO-1st" (\ref{holo:pmn4}, \ref{holo:vassel}), and reconstructed image of hologram at distance $z=8 cm$ (\ref{holo:PMN4-recon}, \ref{holo:vassel-recon}). $x$ and $y$ axis are reported in $px$ (pixel) with $0.5 cm$ between every pixel.}
\vspace{-5mm}
\end{figure}

\subsection{Error calculation}\label{subsubsec:error}

At this point, after the introduction of the simulation examples with Optical Holography (Fig. \ref{fig:holo-lossless}) and with the settings corresponding to ``UGO-1st`` (Fig. \ref{fig:holo-lossy}), we want to calculate and visualize the errors for a limited RADAR synthetic aperture. In this paragraph, we first define two losses, and then calculating the errors with a varying synthetic detector size. Finally, we calculate the errors from real PMN4 landmine hologram taken with HSR (Holographic Subsurface RADAR) \ref{holo:pmn4} and an Archaeological terracotta vessel in Fig. \ref{holo:vassel}. In these two last experiments, we consider the objects at distance of $8$\;$cm$ (defined by the experimental setup of the real configuration in which the top of PMN-4 is located at $5$\;$cm$ distance from scanning plane and its center is at $3$\;$cm$ depth in the object, as in Fig. \ref{fig:ugo-first}).

To estimate the reconstruction error using Angular Spectrum algorithm, implemented by the HoloPy tool \cite{holopy}, we calculate two losses. Both the losses are scaled based on the number of pixels that constitute the reconstructed image (indicated with an $\star$). The first one is the classical $L2$ loss:

\begin{equation*}
    L2^{\star}_{loss} = \frac{1}{n}\sqrt{\sum_{i=1}^n (orig_i - rec_i)^2}
\end{equation*}
 
The second loss is a $L2$ variation which consider only the part of the reconstructed image that should be zeros. To do so, we have calculated a filter that contains a sphere in the middle, and we use that to calculate the loss on out-sphere values:

\begin{equation*}
    L2^{\star}_{loss\_zero} =  \frac{1}{|| \mathbb{1} ||} \sqrt{\sum_{i=1}^n (orig_i - rec_i)^2 \cdot \mathbb{1}_{orig_i =0}}
\end{equation*}
where $\mathbb{1}_{orig_i =0}$ is an indicator function that equals 1 if the value of the original image at index $i$ is 0, and 0 otherwise.

\begin{figure}
\centering
\begin{subfigure}{.25\textwidth}
    \centering
    \includegraphics[scale=0.3]{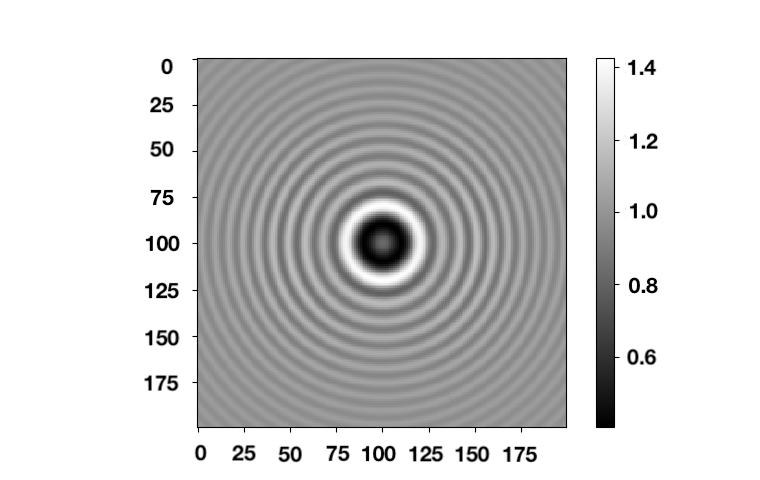}
    \caption{Simulated Hologram image.}
    \label{holo:sphere-real}
\end{subfigure}%
\begin{subfigure}{.25\textwidth}
    \centering
    \includegraphics[scale=0.3]{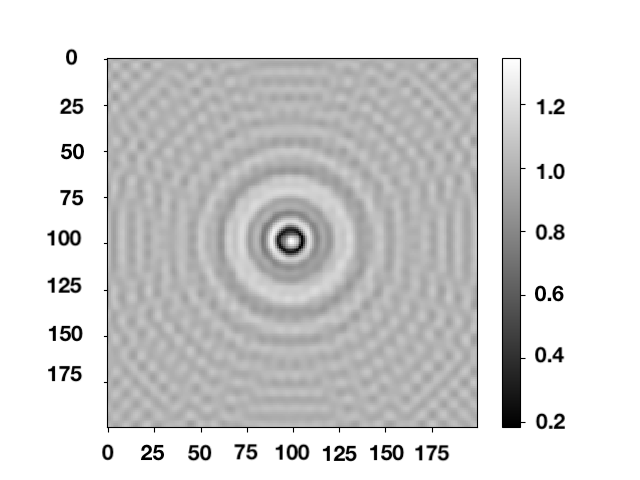}
    \caption{Reconstructed image.}
    \label{holo:sphere-real-recon}
\end{subfigure}
\caption{Simulated Hologram with $200 \times 200$ pixels hologram surface size.}
\label{holo:bigger}
\vspace{-5mm}
\end{figure}

\section{Results}

\begin{figure}
    \centering
    \includegraphics[scale=0.4]{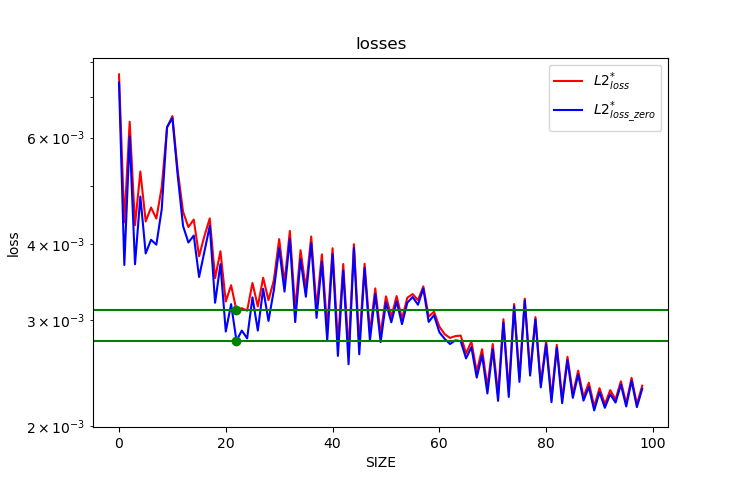}
    \caption{Losses pattern for Sphere simulation error with real HSR configurations. SIZE refers to number of pixels (pixel size: 5 mm) added to each dimension of the original scanner. }
    \label{fig:losses}
\end{figure}

In Fig. \ref{fig:losses} are reported the values of $L2^{\star}_{loss}$ and $L2^{\star}_{loss\_zero}$ errors calculated between the reconstructed image of the sphere and its original image, at different scanning area sizes.

The $x$ axis refer to pixels added to the $52\times62$ detector size, with $0.5$\;$cm$ distance within each pixel. The $y$ axis is reported in logaritmic scale. 
As we can notice from the two lines, the overall trends is decreasing. This indicates that an increase in the detector dimension of $50$\;$cm$ per-side can lower the error of $71.7 \%$ (resulting in an Holographic image of shape $102\;cm$$\times$$112\;cm$). Unfortunately, this detector size is unfeasible for the robot "UGO-1st", making this accuracy gain not possible. However, as the trend of the errors is decreasing, we can expect an accuracy gain from every size change applied to the detector. From the simulations results, we can notice that around the value $x=22$ there is a local minimum of both the losses, that is lowered only after $x>40$. This means that a detector dimension of $63\times73$\;$(cm\times cm)$ could help the image reconstruction with an accuracy increase of $63.7 \%$ (from $7,6$ to $2,76$ in $L2^{\star}_{loss\_zero}$). 

\begin{table}[]
\caption{Errors with $L2^{\star}_{loss}$ and $L2^{\star}_{loss\_zero}$,. Results are reported in magnitude $10^{-2}$.}
\centering
\begin{tabular}{lccccc}
\hline
\textbf{OBJECT} & \textbf{SRC} & \textbf{SIZE (cm)} & \textbf{TYPE ($\lambda;\;freq.$)} & \textbf{$L2^{\star}_{l}$} & \textbf{$L2^{\star}_{lz}$} \\ \hline
PMN4   & real & 52x62 & $15;\;1.9GHz$ & $1.182$ & $1.208$ \\
Vessel & real & 52x62 & $15;\;1.9GHz$ & $1.012$ & $1.066$ \\ \hline
\end{tabular}
\label{tab:results}
\end{table}

In Table \ref{tab:results} are reported the error values of two (real) hologram image reconstruction using the angular spectrum algorithm. The first row refer to the reconstruction of PMN4 landmine, while the second row is related to the Archaeological Terracotta vessel (Fig. \ref{holo:vassel}). The intent of the table is to show the magnitude of real-case errors compared to the simulated settings. In fact, real case errors are bigger than simulated errors by a factor of $10$.

\section{Conclusion}
This paper aims to investigate the influence of the dimensions of the scanned area with microwave HSR on the reconstructed hologram with angular spectrum algorithm. This size is chosen so that it (i) minimizes the error in holographic image reconstruction and (ii) is a size feasible for the demining robot. From simulations, we empirically demonstrated that an increase of $11$\;$cm$ in both $x$ and $y$ dimension of the detector could decrease the reconstruction error of about $63.7 \%$. Additionally, this size bring an accuracy gain comparable with the one we can have by increasing the detector size of more than $21$\;$cm$. In conclusion, this article shows how in extreme holographic acquisition conditions, with wavelengths comparable with the size of the object and the size of the synthetic aperture of the radar, the holographic inversion presents some critical issues. The error estimation proposed in this article offers a tool to evaluate the effects of a limited synthetic aperture of the radar in relation to the wavelength used (2 GHz). This paper suggest to enlarge the scanned area at the cost of more weight and volume of the scanner carried by the robotic platform.

\section*{Acknowledgment}

This research work is founded thanks to ASMARA (by Tuscany Region) and NATO SPS G-5731 projects.

\printbibliography

@ARTICLE{microwave:holo,
  author={Tricoles, G. and Farhat, N.H.},
  journal={Proceedings of the IEEE}, 
  title={Microwave holography: Applications and techniques}, 
  year={1977}
}

@article{acustic:holo,
	title = {Holograms for acoustics},
	volume = {537},
	issn = {0028-0836, 1476-4687},
	doi = {10.1038/nature19755},
	pages = {518--522},
	number = {7621},
	journaltitle = {Nature},
	shortjournal = {Nature},
	author = {Melde, Kai and Mark, Andrew G. and Qiu, Tian and Fischer, Peer},
	date = {2016-09-22},
	langid = {english},
}

@article{gabor:1948,
	title = {A New Microscopic Principle},
	volume = {161},
	doi = {10.1038/161777a0},
	number = {4098},
	journaltitle = {Nature},
	shortjournal = {Nature},
	author = {Gabor, D.},
	date = {1948-05-15},
	langid = {english},
}

@book{holo:book,
  author = {Georges T., Nehmetallah and Rola, Aylo and Logan, Williams},
  year = {2015},
  title = {Analog and Digital Holography with MATLAB},
  doi = {10.1117/3.2190844},
}

@ARTICLE{134436,
  author={Soumekh, M.},
  journal={IEEE Transactions on Signal Processing}, 
  title={Bistatic synthetic aperture radar inversion with application in dynamic object imaging}, 
  year={1991},
  doi={10.1109/78.134436}}

@ARTICLE{128031,
  author={Soumekh, M.},
  journal={IEEE Transactions on Image Processing}, 
  title={A system model and inversion for synthetic aperture radar imaging}, 
  year={1992},
  doi={10.1109/83.128031}}

@article{942570,
title = {Three-dimensional millimeter-wave imaging for concealed weapon detection},
volume = {49},
doi = {10.1109/22.942570},
number = {9},
journaltitle = {{IEEE} Transaction on Microwave Theory and Techniques},
author = {Sheen, D.M. and {McMakin}, D.L. and Hall, T.E.},
urldate = {2020-09-09},
date = {2001},
langid = {english},
}

@InProceedings{10.1007/978-1-4615-8210-6_18,
author="Boyer, A. L.
and Hirsch, P. M.
and Jordan, J. A.
and Lesem, L. B.
and Van Rooy, D. L.",
title="Reconstruction of Ultrasonic Images by Backward Propagation",
booktitle="Acoust. Holography",
year="1971",
}

@Article{electronics11182883,
AUTHOR = {Bossi, Luca and Falorni, Pierluigi and Capineri, Lorenzo},
TITLE = {Versatile Electronics for Microwave Holographic RADAR Based on Software Defined Radio Technology},
JOURNAL = {Electronics},
YEAR = {2022},
DOI = {10.3390/electronics11182883}
}

@software{delafuente,
    author = {De la Fuente, Rafael},
    doi = {10.5281/zenodo.6843673},
    title = {{diffractsim: A flexible python diffraction simulator}},
}

@proceedings {holopy,
	title = {Digital Holographic Microscopy for 3D Imaging of Complex Fluids and Biological Systems},
	journal = {Frontiers of Engineering: Reports on Leading-edge Engineering from the 2009 Symposium},
	year = {2010},
	note = {Full Text Download PDF},
	address = {Irvine, CA},
	author = {Vinothan N. Manoharan}
}

@misc{nato:demining-robot,
	title = {Demining Robots},
	language = {en},
	urldate = {2022-05-31},
	journal = {DEMINING ROBOTS - NATO SPS G-5731},
	url = {www.natospsdeminingrobots.com},
}

@misc{nato:ugo-1st,
title = {Holographic and Impulse Subsurface Radar for Landmine and IED Detection},
url = {http://www.natospsdeminingrobots.com/},
journal = {NATO SPS G-5014}
}

@INPROCEEDINGS{6254884,
  author={Razevig, Vladimir and Ivashov, Sergey and Vasiliev, Igor and Zhuravlev, Andrey},
  booktitle={2012 14th ICGPR}, 
  title={Comparison of different methods for reconstruction of microwave holograms recorded by the subsurface radar}, 
  year={2012},
  volume={},
  number={},
  doi={10.1109/ICGPR.2012.6254884}}

@INPROCEEDINGS{9138227,
  author={Bossi, Luca and Falorni, Pierluigi and Pochanin, Gennadiy and Bechtel, Timothy and Sinton, Jack and Crawford, Fronefield and Ogurtsova, Tetiana and Ruban, Vadym and Capineri, Lorenzo},
  booktitle={IEEE International Workshop on Metrology for Industry 4.0 \& IoT}, 
  title={Design of a robotic platform for landmine detection based on Industry 4.0 paradigm with data sensors integration}, 
  year={2020},
  doi={10.1109/MetroInd4.0IoT48571.2020.9138227}}

\end{document}